\documentclass[reprint, secnumarabic, amssymb, amsmath, superscriptaddress, aps, prl]{revtex4-1}
\usepackage{graphicx}
\usepackage[colorlinks=true,urlcolor=blue]{hyperref}
\usepackage{hyperref,ulem}
\usepackage{xcolor}
\hypersetup{
colorlinks,
linkcolor={red!50!black},
citecolor={blue!50!black},
urlcolor={blue!80!black}
}
\begin{document}

\title{Search for $Q \sim 0$ order near a forbidden Bragg position in Bi$_{2.1}$Sr$_{1.9}$CaCu$_2$O$_{8+x}$ with resonant soft x-ray scattering
}

\author{Xuefei Guo}
\email{xuefeig2@illinois.edu}
\author{Sangjun Lee}
\author{Thomas A. Johnson}
\author{Jin Chen}
\author{Paul Vandeventer}
\author{Ali A. Husain}
\affiliation{Department of Physics and Materials Research Laboratory, University of Illinois, Urbana, Illinois 61801, USA}
\author{Fanny Rodolakis}
\author{Jessica L. McChesney}
\affiliation{Advanced Photon Source, Argonne National Laboratory, Argonne, IL 60439, USA}
\author{Padraic Shafer}
\affiliation{Advanced Light Source, Lawrence Berkeley National Laboratory, Berkeley, CA 94720, USA}
\author{Hai Huang}
\author{Jun-Sik Lee}
\affiliation{Stanford Synchrotron Radiation Lightsource, SLAC National Accelerator Laboratory, Menlo Park,CA 94025, USA}
\author{John Schneeloch}
\author{Ruidan Zhong}
\author{G. D. Gu}
\affiliation{Condensed Matter Physics and Materials Science Department, Brookhaven National Laboratory, Upton, NY 11973, USA}
\author{Matteo Mitrano}
\affiliation{Department of Physics, Harvard University, Cambridge, MA 02138, USA}
\author{Peter Abbamonte}
\email{abbamonte@mrl.illinois.edu}
\affiliation{Department of Physics and Materials Research Laboratory, University of Illinois, Urbana, Illinois 61801, USA}

\begin{abstract}

Identifying what broken symmetries are present in the cuprates has become a major area of research. Many authors have reported evidence for so-called ``$Q \sim 0$" order that involves broken inversion, mirror, chiral, or time-reversal symmetry that is uniform in space. Not all these observations are well understood and new experimental probes are needed. 
Here we use resonant soft x-ray scattering (RSXS) to search for $Q \sim 0$ order in Bi$_{2.1}$Sr$_{1.9}$CaCu$_2$O$_{8+x}$ (Bi-2212) by measuring the region of a forbidden Bragg peak, $(0,0,3)$, which is normally extinguished by symmetry but may become allowed on resonance if valence band order is present. 
Using circularly polarized light, we found that this reflection becomes allowed on the Cu $L_3$ resonance for temperatures $T_c < T < T^\ast$, though remains absent in linear polarization and at other temperatures. This observation suggests the existence of spatially uniform valence band order near the pseudogap temperature. 
In addition, we observed periodic oscillations in the specular reflectivity from the sample surface that resemble thin film interference fringes, though no known film is present. These fringes are highly resonant, appear in all polarizations, and exhibit a period that depends on the location where the beam strikes the sample surface. 
We speculate that these fringes arise from interaction between some intrinsic valence band instability and extrinsic structural surface morphologies of the material. Our study supports the existence of some kind of $Q \sim 0$ broken symmetry state in Bi-2212 at intermediate temperatures, and calls for further study using a microfocused beam that could disentangle microscopic effects from macroscopic heterogeneities. 

\end{abstract}

\maketitle

\section{Introduction}

The origin of superconductivity in copper oxides is still not understood after more than three decades of investigation. Adding to the challenge is that the cuprates are subject to a host of spin, charge, and nematic instabilities whose spectroscopic signatures are difficult to identify and disentangle from those of superconductivity itself \cite{keimer2015quantum}. Establishing what other order parameters may be present, and whether they are related to important aspects of cuprate phenomenology (e.g., the pseudogap) is therefore a major focus of current research. 

Many recent studies have focused on instabilities that are periodic or quasiperiodic in real space, such as spin density waves, charge density waves, or more exotic intertwined states such as the proposed pair density wave (PDW) \cite{berg2009a,berg2009b}. However, an important sub-category of candidate orders are those that occur at $Q \sim 0$, i.e., phenomena that are uniform in space and have signatures in the average spectroscopic properties of the material. Many experimental studies have reported evidence for such order interpreted in terms of broken time reversal symmetry \cite{kaminski2002spontaneous,he2011single,xia2008polar,fauque2006magnetic,mook2008observation,
baledent2011evidence,de2012evidence,mangin2014characterization,mangin2015intra,mangin2017b,
tang2018orientation,li2008unusual,li2011magnetic,croft2017no,zhao2017global}, inversion symmetry \cite{zhao2017global,lim2020observation}, rotational symmetry \cite{hinkov2008electronic,daou2010broken,sato2017thermodynamic,achkar2016nematicity,
lawler2010intra,mesaros2011topological,fujita2014simultaneous,mukhopadhyay2019evidence,
ishida2020divergent}, chiral symmetry \cite{lim2020observation}, mirror symmetry\cite{delatorre2021}, as well as anapole symmetry that may be related to proposals of ``loop current" order \cite{kaminski2002spontaneous,fauque2006magnetic,mook2008observation,baledent2011evidence,
de2012evidence,mangin2014characterization,mangin2015intra,mangin2017b,tang2018orientation,
li2008unusual,li2011magnetic,croft2017no,scagnoli2011observation}.
However, some of these experiments still remain controversial and new experimental schemes capable of detecting exotic valence band order near $Q \sim 0$ are needed. 

Resonant soft x-ray scattering (RSXS) is a momentum-resolved spectroscopic technique that is capable of coupling to a variety of ordered valence band states, in principle those comprising higher order electric and magnetic multipoles \cite{fink2013resonant,lovesey2005}. The technique is sensitive to the symmetry of valence band order through the structure of the scattering tensor that arises when the incident photon energy is tuned to atomic resonances such as the Cu $L_3$ edge. Signatures of $Q \sim 0$ order appear in RSXS in two ways, either through the resonance behavior of Bragg reflections, which represent the behavior of phenomena with the same periodicity as the structural lattice, or through the average optical properties of the material detected in specular reflectivity from the material surface. 

Bi$_{2.1}$Sr$_{1.9}$CaCu$_2$O$_{8+x}$ (Bi-2212), the material that we focus on in this study, has a base-centered orthorhombic average crystal structure \cite{di2007x}. Some Bragg reflections in this material---for example, the $(0,0,3)$ reflection---are extinguished by symmetry due to the equivalence of the Cu atoms in different layers. This extinction arises because of the approximate spherical symmetry of the Cu atomic density in the scattering regime dominated by Thomson scattering. However, near resonance where scattering is tensorial, this extinction is no longer guaranteed if local multipole moments are coherently arranged in different layers \cite{murakami1998}. Measuring the RSXS signal near forbidden $(0,0,L)$ reflections on resonance, where $L$ is odd, is therefore a highly sensitive way to potentially detect $Q \sim 0$ order, in principle even if it involves anapole or other exotic symmetries. Such an approach yields extremely high sensitivity, since it is effectively a difference measurement between two nearly equivalent sites, which as tensor objects are less likely to cancel as they do in simple Thomson scattering \cite{murakami1998}. Partial information about the symmetry of the scattering tensors can be attained by measuring the polarization dependence of the scattering. 

\begin{figure*}
\includegraphics[width=0.8\linewidth]{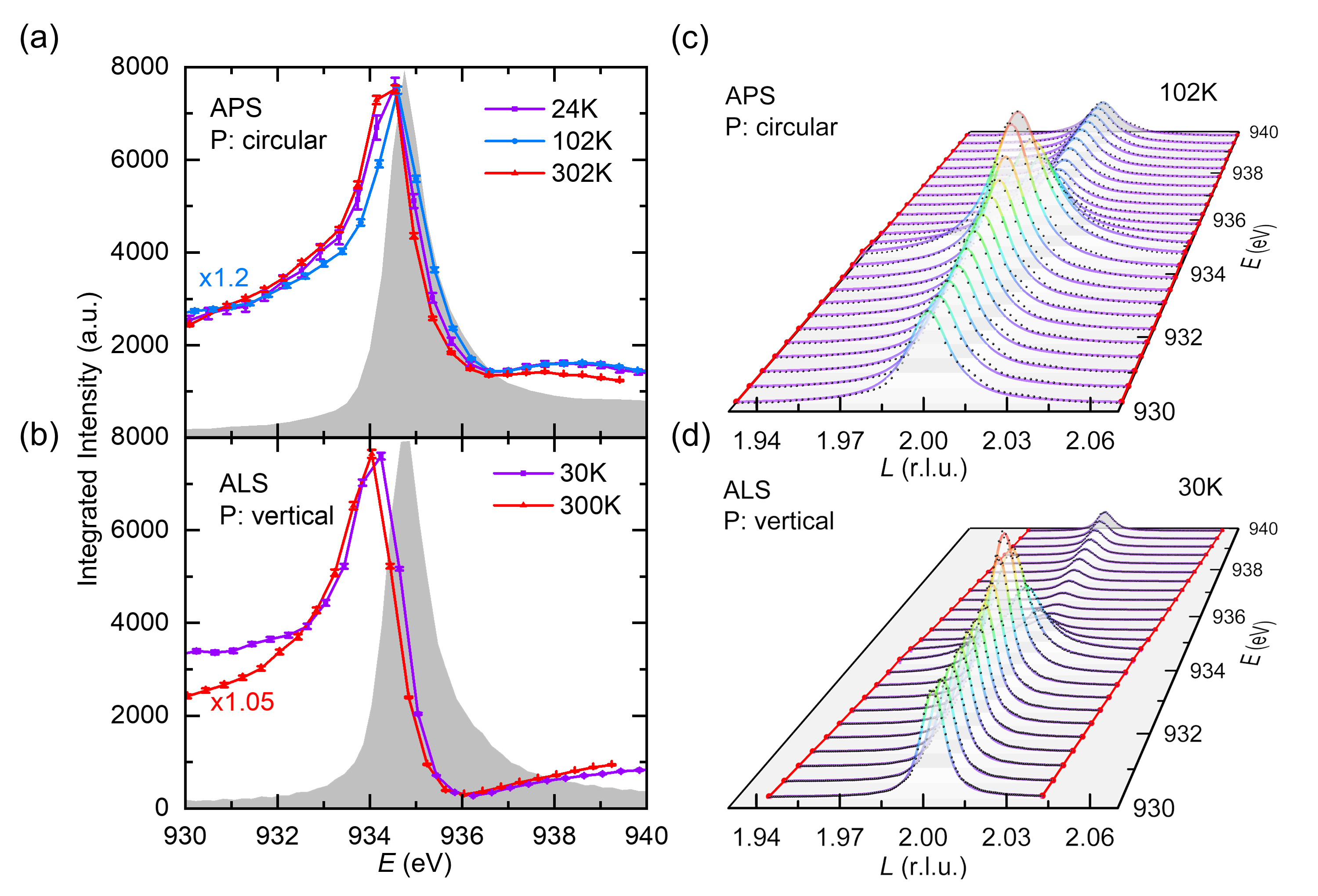}
\caption{\label{fig1}Resonance profiles of the structural $(0,0,2)$ Bragg peak. (a) Temperature dependence of the $(0,0,2)$ intensity measured with with circularly polarized light at APS. The integrated intensity was obtained by fitting the  peak with a Lorentzian. The XAS spectrum (gray background) is shown for comparison. (b) Temperature dependence of the $(0,0,2)$ intensity using linearly polarized light at ALS. The integrated intensity was obtained from Voigt fits. The XAS (gray background) is, again, shown for reference. (c) Representative momentum scans with taken at $T=102$ K with circularly polarized light, with individual data points shown in black and the fit curves shown as colored lines. (d) Representative, individual momentum scans with taken at $T=30$ K with linearly polarized light using the same color scheme as in panel (c). 
}
\end{figure*}

Here, we present RSXS measurements of the forbidden $(0,0,3)$ reflection on the Cu $L_3$ edge in Bi-2212 above and below $T^\ast$. Intriguing oscillations of the scattering intensity and resonance features of forbidden and specular reflections, that change with both temperature and polarization, are observed. These intriguing results should influence future thinking about what competing orders may be present in the cuprates. 

\section{Experiment}

Single crystals of bilayer Bi-2212 were grown using floating zone methods described previously \cite{wen2008}. 
An optimally doped crystal with $T_c=91$ K and $T^\ast\sim200$ K was chosen because $T_c$ and  $T^\ast$ are split and accessible with our experiment. The sample surface was prepared by cleaving in air using Scotch tape and immediately transferring the sample into the vacuum system. 

RSXS measurements were performed near the Cu $L_{3}$ edge as in many past studies \cite{abbamonte2004crystallization, abbamonte2005spatially, damascelli2014charge, comin2015broken, ghiringhelli2012long, comin2016resonant, frano2020charge}. Experiments were carried out at the Advanced Light Source (ALS) beamline 4.0.2 and the Advanced Photon Source (APS) Sector 29. 
The incident beam at APS has circular polarization while the beam at ALS has vertical polarization, with both setups employing polarization-integrating detectors. Comparison between the two facilities therefore provides partial, though not complete, information about the symmetry of the scattering tensor, which could be deduced through appropriate modeling that we will not undertake here.

Otherwise, the setups at the two beamlines were comparable. We performed Cu $L_3$ edge x-ray absorption (XAS) measurements in total electron yield mode at APS and total fluorescence yield at ALS (Fig. 1(a),(b)), and then used the measured edge position to correct for slight differences in the beam energy between the two facilities. The degree of second harmonic contamination near the Cu $L_3$ edge, i.e., 1870 eV photons in the nominally 935 eV beam, was also nearly the same---about 1-2\% of the fundamental intensity at both facilities. Both setups employed energy-integrated detectors, without an energy analyzer. The APS setup used a photodiode detector with an angular resolution of 3$^{\circ}$. The ALS setup used the same type of detector, but with an angular resolution of 0.15$^{\circ}$. In addition, a charge-coupled device (CCD) detector was available at ALS, as discussed further below. All measurements were carried out using a photodiode unless otherwise specified. 

In this study we denote momenta using orthorhombic units, i.e., Miller indices $(H,K,L)$ designate a momentum transfer $Q=2 \pi (H/a,K/b,L/c)$, where $a = b = 5.4 \AA$ and $c=30.9 \AA$. The sample was oriented so that the $(1,1,0)$ direction, which is $45^{\circ}$ from the structural  supermodulation, was placed in the horizontal scattering plane. The intensity units were calibrated between the two facilities by matching the $(0,0,2)$ peak intensity on resonance at 24 K at APS and at 30 K at ALS, compensating for differences in flux and detector efficiency between the two facilities. 

RSXS instruments are highly suitable for detecting $Q\sim0$ order, as they are capable of measuring both Bragg peaks and specular reflectivity. The former appear as sharp peaks at integer values of $(H,K,L)$ while the latter forms a continuous rod along the $(0,0,L)$ direction, as we discuss further below. 

\section{Results}

\begin{figure*}
\includegraphics[width=1\linewidth]{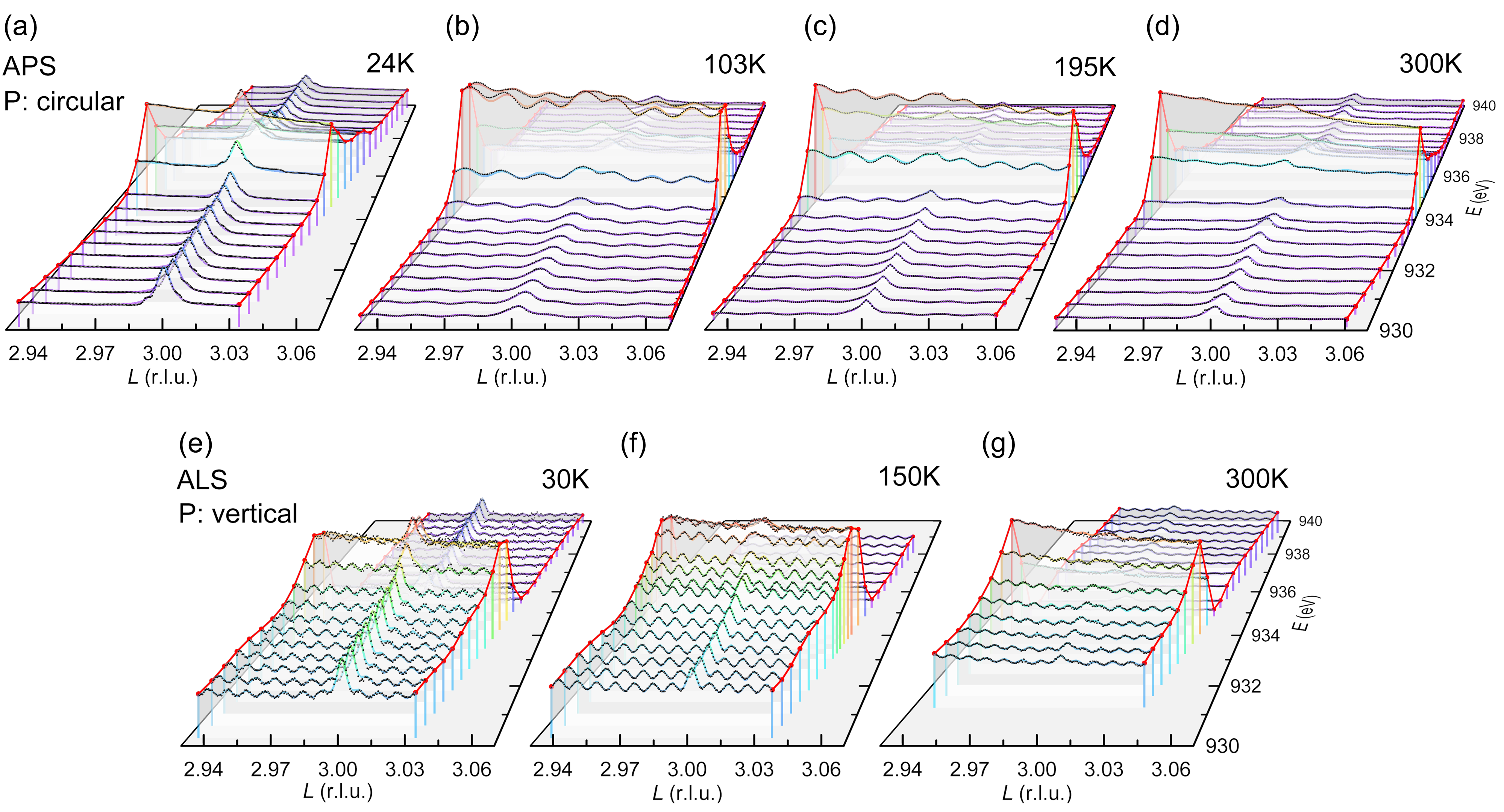}
\caption{\label{fig2}RSXS momentum scans in the vicinity of the forbidden Bragg peak, $(0,0,3)$, for two different polarizations, a variety of temperatures, and a variety of different incident energies across the Cu $L_3$ edge. The black points are the experimental data and the colored lines represent fits using Eq. 1. (a)-(d) Measurements using circular polarization at APS. (e)-(g) Measurements using linear polarization at ALS.  
}
\end{figure*}

We started by performing a routine measurement of the $(0,0,2)$ Bragg peak. This is a normal, allowed structural Bragg reflection expected to show no changes with temperature. Therefore, such measurements are useful for validating our subsequent measurements on forbidden Bragg peaks and specular reflections, and also serve as an opportunity to compare performances between the two facilities.  

The results are summarized in Fig. \ref{fig1}. We performed momentum scans through the $(0,0,2)$ reflection at a variety of incident energies across the Cu $L_3$ edge at several temperatures, as shown in the line curves in Fig. \ref{fig1}(c)-(d). The results are summarized as resonance profiles in Figs. \ref{fig1}(a)-(b), which show the integrated intensity of the $(0,0,2)$ Bragg peak as a function of incident energy, compared to the XAS spectrum. The profile forms an asymmetric peak that remains strong off the resonance visible at both facilities, which arises from interference between resonant scattering and non-resonant Thomson scattering. This shape is typical of anomalous scattering from a structural reflection whose existence is not tied to subtle details in the electronic structure \cite{abbamonte2006charge}. Figs. \ref{fig1}(a)-(b) show that, due to the drifts in the beam current and other instabilities in the beamline, the overall intensity could drift by as much as 20\% during a weeks-long experimental run (see in particular Fig. \ref{fig1}(a)). Apart from overall intensity drifts, the results at the two facilities are highly consistent. This confirms that the configurations of the two beamlines are the same and that we can compare the results within an absolute intensity of 20\% (one caveat is that the harmonic contribution may be more unstable than the fundamental energy, as discussed further below). No temperature dependence of the $(0,0,2)$ reflection was observed when cooling through  $T^\ast$, confirming expectations that the overall crystal structure does not change at $T^\ast$.

Having validated the setups, we performed the same measurements near the $(0,0,3)$ point in momentum space. This reflection is forbidden by the crystal structure so, to a first approximation, no Bragg scattering is expected, apart from a weak second harmonic background from the $(0,0,6)$ reflection diffracting $\sim$1870 eV photons. Note that harmonic scattering can be identified because it is energy-independent, since there is no resonance in Bi-2212 at around 1870 eV. However, as discussed in the introduction, a true $(0,0,3)$ reflection may become weakly allowed if there is some symmetry breaking in the valence band that makes Cu atoms in two of the layers inequivalent. 

Measurements near $(0,0,3)$ are summarized in Fig. \ref{fig2}, which shows individual momentum scans through the $(0,0,3)$ location at a variety of incident energies across the Cu $L_3$ edge at several temperatures between 24 K and 300 K. Identical scans were done at APS and ALS, i.e., with beams having circular and vertical polarization, respectively.

Starting with circular polarization at 24 K (Fig. \ref{fig2}(a)), the spectrum shows a weak reflection at $L=3$ sitting on a strong specular reflectivity background that is $L$-independent. The specular intensity is strongly energy-dependent, showing an enhancement at the Cu $ L_3$ edge, which is expected because the optical constants of the material are strongly energy-dependent. The peak near the $(0,0,3)$ position, however, was found to be energy-independent through the Cu $L_3$ edge. This observation is consistent with pure harmonic scattering, i.e., diffraction of 1870 eV background light from the $(0,0,6)$ reflection. This measurement provides a useful measure of the degree of harmonic contamination in the beam.

At an increased temperature of 103 K, three new features appear in the data, as shown in Fig. \ref{fig2}(b). First, periodic oscillations are observed with a period of $\delta L = 0.02$ r.l.u. (reciprocal lattice units). These oscillations resemble interference fringes from a thin film with a thickness of 150 nm, though the sample has a thickness of 100 $\mu$m, was freshly cleaved, and no known layer was present. Also, the fringes are not discernible at 24 K. 
These oscillations are highly energy-dependent near the Cu $L_3$ edge, as shown in Fig. \ref{fig2}(b), identifying them as arising from the Bi-2212 crystal and not an extrinsic overlayer. Second, the intensity of the $L$-independent specular reflectivity increases by a factor of three near the Cu $L_3$ resonance, compared to the intensity at 24 K. This enhancement was not observed in the off-resonant signal and appears to signify a change in the $Q=0$ optical constants near the Cu $L_3$ resonance. This change would be expected if there were some form of $Q=0$ valence band order in the material. Third, a true $(0,0,3)$ Bragg feature appears that was not obviously discernible at 24 K. The peak interferes with the oscillations, and appears in the data as a distortion to the fringes near the $L=3$ location (Fig. 2(b)). This distortion is strongly energy-dependent, indicating that this is a true $(0,0,3)$ reflection has acquired a resonant component not observed at 24 K.

At $T=$ 195 K, as shown in Fig. \ref{fig2}(c), the fringes and the specular reflectivity are still strong on resonance, with intensity similar to 103 K. The (003) feature is also still present, though is somewhat weaker than at 103 K. 

Finally, at room temperature (Fig. 2(d)), the fringes are no longer clearly visible. The specular reflectivity is significantly reduced, though it is still stronger than at 24 K. The $(0,0,3)$ reflection is now absent again. The remaining Bragg feature is, again, an energy-independent harmonic signal from the $(0,0,6)$ reflection, similar to what was observed at 24 K. 

These results are surprising. There is no clear reason for the presence of interference oscillations, since the measurements were performed on a freshly cleaved single crystal on which no known thin film is present. Further, the changes in the specular rod and the $(0,0,3)$ reflection at intermediate temperatures suggest some alteration of the $Q=0$ optical properties of the material in the temperature range 100 K $<T<$ 200 K, i.e., between $T_c$ and $T^\ast$. 
Moreover, the $(0,0,3)$ reflection is only visible in this same temperature range and only on resonance, suggesting it could indicate some kind of $Q \sim 0$ electronic ordering in the material. 

To assess the reproducibility of these results, and to explore how they depend on the beam polarization, we repeated the experiment at ALS using linearly polarized light. We used the same sample as at APS, but we cleaved it again to achieve a fresh surface and remove any possible thin film that may have been present. 

At $T=30$ K, as before, a weak but clear peak near $(0,0,3)$ was observed (Fig. \ref{fig2}(e)). As at APS, the peak is energy-independent, suggesting it is again coming from the $(0,0,6)$ reflection diffracting the harmonic content in the beam. 
The intensity is similar to that at APS, confirming expectations that the harmonic content at the two facilities is similar. 
Despite the new cleave, clear oscillations were again observed, this time with a period $\delta L = 0.0085$ r.l.u., that resemble interference fringes from a film with thickness 360 nm. The oscillations were strongly energy-dependent near the Cu $L_3$ edge demonstrating, again, that they are a property of the Bi-2212 crystal, though the periodicitiy differs from what was observed with circularly polarized light by more than a factor of two. 
Both the $(0,0,3)$ peak and the oscillations sit on a specular reflectivity background that is strongly energy dependent. Strangely, the energy dependence of the fringes and the specular reflectivity are both very different from what was observed using circularly polarized x-rays at APS. We will summarize this difference quantitatively below.

Increasing the temperature to 150 K (Fig. \ref{fig2}(f)), the spectrum changes very little. The peak near the $(0,0,3)$ position is still present and energy-independent, indicating that it is purely due to harmonic scattering. However, the intensity is lower than at 30 K. This change is most likely due to changes in the harmonic content in the beam, which is more subject to subtle alignment drifts than the fundamental. Oscillations are still present with the same intensity and period as at 30 K. This behavior is very different from what was observed with circular polarization at APS, where the oscillations are barely visible at 24 K but are strong at 103 K. The reflectivity has the same intensity and energy dependence as at 30 K. 

Warming to 300 K, the peak near $(0,0,3)$ still energy-independent (Fig. \ref{fig2}(g)), indicating that it arises solely from harmonic scattering at all temperatures. In other words, a true $(0,0,3)$ reflection is not observed at any temperature with linearly polarized x-rays. The oscillations are still present at this temperature but are weaker than at 150 K by a factor of two. The specular reflectivity is unchanged, and is essentially temperature-independent in this polarization.

\begin{figure*}
\includegraphics[width=1\linewidth]{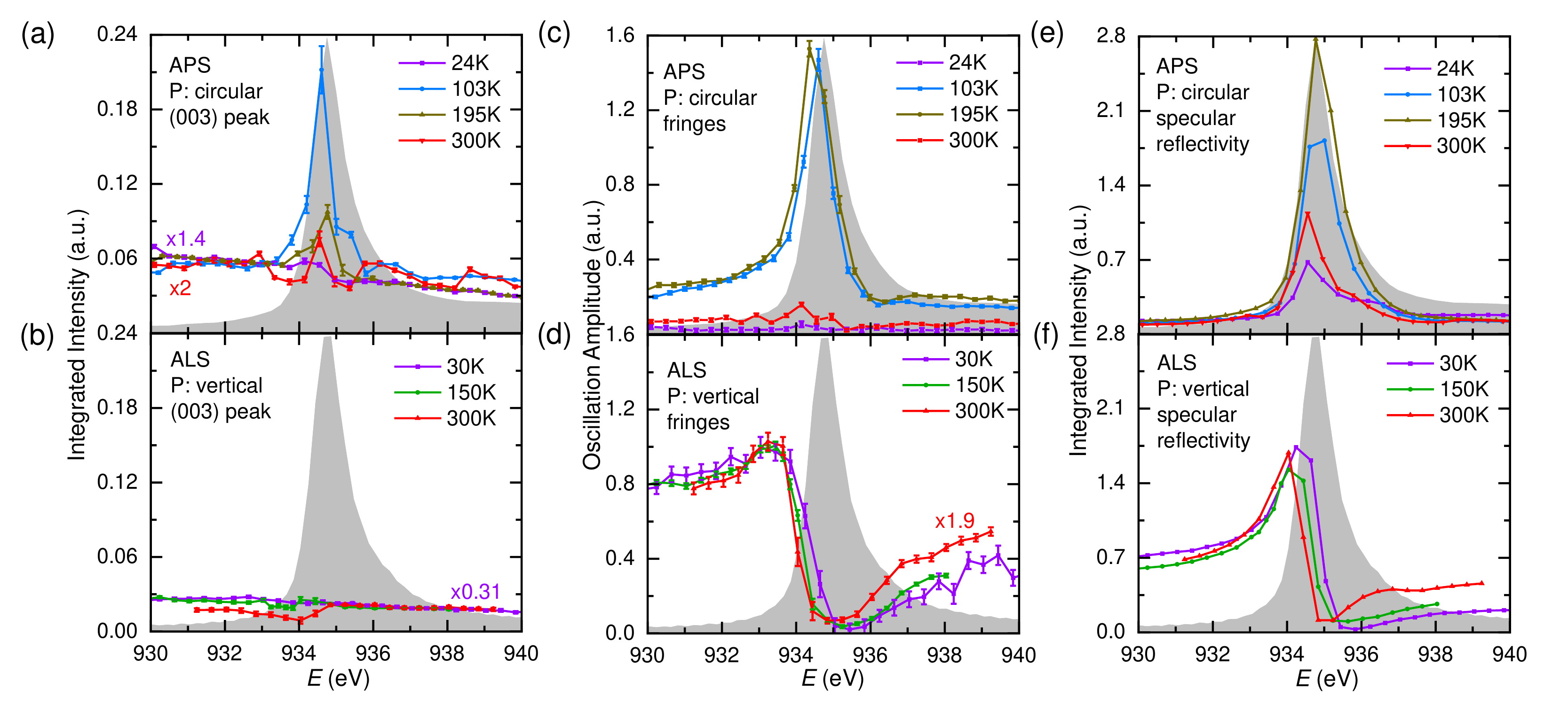}
\caption{\label{fig3}
Resonance profiles of the primary features in the RSXS measurements shown in Fig. 2. 
(a) Integrated intensity of the $(0,0,3)$ peak plotted as a function of incident beam energy taken at several temperatures with circularly polarized light at APS. (b) Same plot as panel (a) but measured with linear polarization at ALS. (c) Amplitude of the interference fringes plotted as a function of incident beam energy at several temperatures with circularly polarized light. (d) Same plot as panel (c) but measured with linear polarization. (e) Integrated intensity of the parabolic, specular background plotted as a function of incident beam energy taken at several temperatures with circularly polarized light. (f) Same plot as panel (e) but measured with linear polarization. The XAS spectrum (gray background) is shown in all panels for comparison.
}
\end{figure*}

\section{Summary Fits}

A summary of all results near the $(0,0,3)$ location in momentum space using both polarizations is shown in Fig. \ref{fig3}. To create this figure, we fit all the individual curves in Fig. \ref{fig2} using a fit function consisting of a Lorentzian for the Bragg peak, a sinusoidal function for the oscillations, and a parabola for the specular component,

\begin{eqnarray}
I\left( q_L \right) = \frac{A_I}{\pi}\frac{\sigma}{(q_L-q_c)^2 + \sigma^2} 
+ O_s\sin\frac{2\pi(q_L-q_0)}{\Delta q} \\ \nonumber
+ aq_L^2 + bq_L+c.
\label{fit_function}
\end{eqnarray}


\noindent
In this expression, $A_I$ represents the integrated intensity of the Bragg peak, $\sigma$ is its width, $O_s$ is the amplitude of the oscillations, and $\Delta q$ represents the oscillation period. This fit function allows us to decompose the experimental data into different components so they can be examined separately. 

Figs. \ref{fig3}(a)-(b) show the $(0,0,3)$ Bragg peak resonance profiles, i.e., the integrated intensity of the peak as a function of incident energy compared to the XAS spectrum. For the circularly polarized beam at APS, a resonant $(0,0,3)$ reflection is present only in the temperature range $T_c < T < T^\ast$, suggesting some $Q \sim 0$ ordering in which Cu atoms in different layers become inequivalent, may be present in the vicinity of the pseudogap temperature. This signature is not present at room temperature or below $T_c$, suggesting it may be somehow connected to the 
pseudogap state. For the linearly polarized beam at ALS, however, this resonant feature is not present at any temperature, suggesting that the $Q\sim0$ order exhibits some multipolar symmetry that couples only to circularly polarized light. 

The same kind of resonance plots for the interference fringes is shown in Figs. \ref{fig3}(c)-(d), which show the amplitude of the oscillations as a function of incident energy, again compared to the XAS spectrum. Oscillations are observed at all temperatures and polarizations. For circularly polarized light, the oscillations are strongest in the temperature range 100 K $<T<$ 200 K when the resonant $(0,0,3)$ reflection is present, but are barely visible at room temperature or below $T_c$. However, for linearly polarized light, the oscillations are present at all temperatures, though the amplitude is lower at room temperature. 

The energy dependence of the oscillations is very different in linear and circular polarization. In
the circular case, the oscillations resonate near the peak of the edge, which is characteristic of valence band order \cite{abbamonte2005spatially, damascelli2014charge, ghiringhelli2012long, fink2011phase, achkar2012distinct, miao2017high, blanco2013momentum, thampy2014rotated, tabis2014charge, wu2012charge, wang2020high,boschini2021dynamic,chen2019charge,blanco2014resonant,he2016observation,jang2017superconductivity,bluschke2019adiabatic,peng2018re,da2015charge,
chaix2017dispersive,he2018persistent,betto2020imprint,forst2014melting}. By contrast, in linear polarization the resonance shows an inflection point characteristic of structural scattering \cite{abbamonte2006charge}, similar to the $(0,0,2)$ reflection shown in Fig. \ref{fig1}. The different energy profiles suggest that whatever is giving rise to these oscillations is a combination of structural and valence band effects, with the former scattering strongly in linear polarization and the latter scattering strongly in circular polarization.

The behavior of the featureless, specular component of the data is summarized in Figs. \ref{fig3}(e)-(f). The behavior is similar to that of the fringes. For circularly polarized light, the reflectivity is strongest in the temperature range $T_c <T< T^\ast$, where the resonant $(0,0,3)$ reflection is present, exhibiting the behavior of valence band scattering. For linearly polarized light, the behavor looks like that of pure structural scattering. The strong similarity between the oscillatory and parabolic contributions suggest that these two effects have a common origin and may be considered as a single phenomenon. The overall behavior of the reflectivity and the fringes points to some form of extrinsic structural inhomogeneity, perhaps related to the detailed surface morphology or the way the material cleaves, interplaying in a nontrivial way with intrinsic electronic instabilities in the material.

\begin{figure*}
\includegraphics[width=0.8\linewidth]{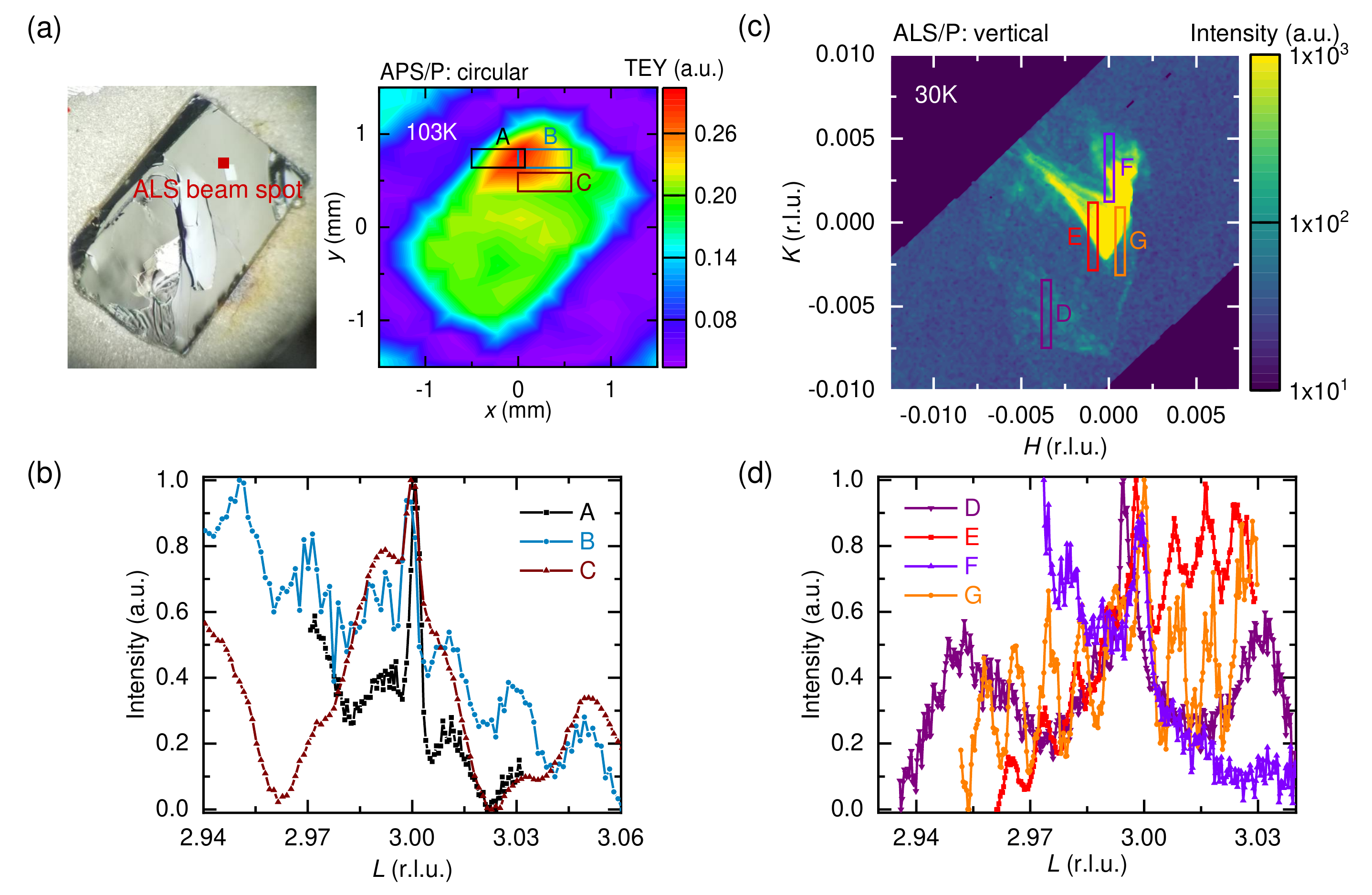}
\caption{\label{fig4}Position and momentum dependence of the oscillations observed in Fig. 2. (a) Microscope image of the cleaved sample after RSXS measurements. The adjacent panel shows a color map of the total electron yield (TEY) at the peak of the Cu $L_3$ edge from the same sample, taken as the beam is scanned across the surface. 
The dark red square in the microscope image shows the position of the beam used for measurements at ALS. The three rectangular boxes show the size and location of the beam spot at APS used for several measurements. 
(b) Momentum scans in the $(0,0,L)$ direction for the three different beam locations, A, B, and C, indicated in panel (a). The overall shape and periodicity of the oscillations strongly depends on the location of the beam on the surface. 
(c) In-plane momentum map at $L = 2.996$ r.l.u. near the Cu $L_3$ edge (6.5 eV off resonance) taken with a CCD detector at ALS. The rectangular boxes, D, E, F, and G, show the ranges of integration in the $(H,K)$ plane for a selection of in-plane momentum values used in panel (d). (d) Momentum scans in the $(0,0,L)$ direction for several different in-plane momenta indicated in panel (c). The oscillations are always visible but their period and overall shape strongly depends on the in-plane momentum chosen. 
}
\end{figure*}

One speculation is that, when the crystal cleaves, it gives rise to some electronic, self-organized layer near the surface of the material whose thickness depends on details of the fracture dynamics or local residual strain. In this case, one might expect the amplitude and period of the fringes to vary significantly on macroscopic scales over the sample surface. We decided to test this possibility by examining the extent to which the oscillations depend on where the x-ray beam strikes the surface.
Using circularly polarized x-rays at APS, we studied the behavior of the fringes at 103 K as the beam is scanned across the sample surface. The results are shown in Figs. \ref{fig4}(a)-(b). 
Fig. \ref{fig4}(a) shows a microscope image of the sample after this cleave next to a map of the total electron yield for the beam striking different locations, which is sensitive to local surface morhphology. We focused our measurements on a visually flat region of the sample surface where the electron yield was high, and then carried out momentum scans in the $(0,0,L)$ direction at several locations denoted A, B and C. The results are summarized in Fig. \ref{fig4}(b). Oscillations are present in all curves, but the periodicity and overall shape is very different for different beam  positions. This observation supports the conclusion that the oscillations, while resonant and seemingly indicating a valence band effect, depend on details of the local surface structure and therefore are also partly an extrinsic phenomenon. 

Another way to view this effect is to examine the off-specular scattering, i.e., the nonzero values of $H$ or $K$ which corresponds to having a small in-plane component of the momentum transfer. Such a measurement would reveal whether the fringes correlate with different in-plane periodicities of the crystal surface.

To obtain information about the off-specular scattering, we repeated the measurement at ALS with linearly polarized light at $T=30$ K, this time using a CCD area detector, which provides a 2D image of near-specular scattering corresponding to small in-plane momenta. The results are shown in Figs. \ref{fig4}(c)-(d). 
Fig. \ref{fig4}(c) shows the $(H,K)$ plane at a fixed momentum $L = 2.996$ r.l.u.. Surprisingly, the results do not show a simple symmetric peak at $H=K=0$. Rather, an intricate structure is observed that involves a significant amount of off-specular scattering. This structure is likely a convolution of two different effects: (1) coherent off-specular scattering due to non-trivial in-plane periodicities, and (2) specular scattering from a surface that is not flat or is corrugated because of morphology of the cleaved surface. Fig. \ref{fig4}(d) shows how the fringes vary with $H$ and $K$, i.e., how the periodicity of the fringes in the $L$ direction depends on the in-plane momentum for several different $(H, K)$ values denoted as D, E, F and G in Fig. \ref{fig4}(c). Again, the fringes are present for all values of $(H,K)$, but the periodicity varies and depends on the chosen in-plane momentum. Evidently, the period of the fringes does not just depend on where the beam hits the sample surface, but also on the form of the surface morphology at any given point. This measurement supports a conclusion that the oscillations are strongly dependent on the detailed local structure of the surface, and so involve some interaction between valence instabilities of the surface and extrinsic effects resulting from the way the crystal cleaves.

\section{Discussion}
To summarize, we performed RSXS measurements near the forbidden Bragg reflection $(0,0,3)$ to explore whether some evidence might be found for $Q=0$ electronic ordering in optimally doped Bi-2212. We found that the $(0,0,3)$ reflection, while forbidden by normal x-ray extinction rules, becomes allowed on resonance when scattering circularly polarized x-rays in the temperature range $T_c < T < T^\ast$. Its resonance properties are typical of valence band order \cite{abbamonte2005spatially, damascelli2014charge, ghiringhelli2012long, fink2011phase, achkar2012distinct, miao2017high, blanco2013momentum, thampy2014rotated, tabis2014charge, wu2012charge, wang2020high,boschini2021dynamic,chen2019charge,blanco2014resonant,he2016observation,jang2017superconductivity,bluschke2019adiabatic,peng2018re,da2015charge,
chaix2017dispersive,he2018persistent,betto2020imprint,forst2014melting}, suggesting some 
$Q \sim 0$ order may be present in this narrow temperature regime near the pseudogap temperature. 
Outside this temperature range, or using linearly polarized x-rays, this reflection remains forbidden. This result might indicate that the order exhibits a multipolar symmetry that couples only to circularly polarized light. Further studies, perhaps using azimuthal rotations or a polarimeter, are needed to reproduce this result and identify the detailed symmetry.

In addition, we observed highly reproducible periodic oscillations that resemble interference fringes from a thin film, though no layered structure is known to be present on the material. The resonance properties of these fringes, in circular polarization, resemble those of the $(0,0,3)$ reflection, suggesting they are a valence band effect. However, in linear polarization these oscillations exhibit the resonance behavior of a structural phenomenon. These fringes are present at all temperatures and polarizations, even after several cleaves of the sample surface. However, their periodicity varies greatly across the sample surface and for different values of the in-plane momentum. We suggest that these fringes are due to some interaction between intrinsic valence band instabilities in the material and extrinsic structural morphologies of the sample surface. 

Concerning the question of $Q=0$ order, our results do not conclusively demonstrate the existence of ordered multipoles or loop currents in the pseudogap regime. However, they demonstrate that some nontrivial valence band ordering effect is taking place and further studies are warranted that could disentangle intrinsic ordering phenomena from effects influenced by the surface properties. An ideal approach would be to perform RSXS studies with a highly focused beam that could investigate a single location on the sample surface down to a sub-micron dimension whose local properties are better defined. This could be accomplished, for example, at the Coherent Soft X-ray (CSX) Scattering facility of the National Synchrotron Light Source II or any other coherent or microbeam instruments currently operational worldwide.

\begin{acknowledgements}
We thank C. M. Varma for inspiring us to undertake this study, and S. L. Cooper for carefully reading the manuscript. This work was supported by the Center for Quantum Sensing and Quantum Materials, an Energy Frontier Research
Center funded by the U. S. Department of Energy, Office of Science, Basic Energy Sciences under Award DE-SC0021238. P. A. gratefully acknowledges support from the EPiQS program of the Gordon and Betty Moore Foundation, Grant No. GBMF9452. Crystal growth was supported by DOE Grant No. DE-SC0012704.

\end{acknowledgements}

\bibliography{Q_0_bib}

\end{document}